\documentclass[aps,twocolumn,prl]{revtex4-1}
\usepackage{upgreek}
\usepackage{amsmath}
\usepackage{graphicx}
\usepackage{ucs} 
\usepackage{CJK}
\usepackage{physics}
\usepackage{ucs} 

\usepackage{color}
\definecolor{darkblue}{rgb}{0,0,0.5}
\definecolor{lila}{rgb}{0.3,0,0.3}
\definecolor{turq}{rgb}{0,0.1,0.4}
\definecolor{lightblue}{rgb}{0.7,0.7,0.9}
\usepackage{url} 

\usepackage[pdftex,
 colorlinks=true,
 backref=page,
 linkcolor=darkblue, 
 filecolor=red,
 citecolor=turq, 
 urlcolor=lila, 
 pdftitle={A narrow-band sodium-resonant fiber-coupled single photon source},
 pdfauthor={Guilherme Stein and Vladislav Bushmakin and Yijun Wang and Andreas W. Schell and Ilja Gerhardt},
 pdfsubject={},
 pdfkeywords={Single Photons, Optical Fibers, Single Molecules, Anti-bunching, Rabi-Oscillations},
 breaklinks=false,
 plainpages=false,
 backref=false,
 bookmarks,
 bookmarksnumbered=true]{hyperref}

\begin{document}
\begin{CJK*}{UTF8}{}  
\title{A narrow-band sodium-resonant fiber-coupled single photon source}

\author{Guilherme Stein}
\affiliation{3.\ Physikalisches Institut, Universit\"at Stuttgart and Stuttgart Research Center of Photonic Engineering (SCoPE), Pfaffenwaldring 57, D-70569 Stuttgart, Germany}
\author{Vladislav Bushmakin}
\affiliation{3.\ Physikalisches Institut, Universit\"at Stuttgart and Stuttgart Research Center of Photonic Engineering (SCoPE), Pfaffenwaldring 57, D-70569 Stuttgart, Germany}
\CJKfamily{gbsn}
\author{Yijun Wang (王奕钧)}
\affiliation{3.\ Physikalisches Institut, Universit\"at Stuttgart and Stuttgart Research Center of Photonic Engineering (SCoPE), Pfaffenwaldring 57, D-70569 Stuttgart, Germany}
\author{Andreas W.\ Schell}
\affiliation{CEITEC -- Central European Institute of Technology, Brno University of Technology, 612 00 Brno, Czech Republic}
\affiliation{Institut f\"ur Festk\"orperphysik, Leibniz Universit\"at Hannover, Appelstra\ss e 2, 30167 Hannover, Germany}
\affiliation{Physikalisch-Technische Bundesanstalt, Bundesallee 100, 38116 Braunschweig, Germany}

\author{Ilja Gerhardt}
\email{i.gerhardt@fkf.mpg.de}
\affiliation{3.\ Physikalisches Institut, Universit\"at Stuttgart and Stuttgart Research Center of Photonic Engineering (SCoPE), Pfaffenwaldring 57, D-70569 Stuttgart, Germany}

\begin{abstract}
Quantum technology requires the creation and control over single photons as an important resource. We present a single photon source based on a single molecule which is attached to the end-facet of an optical fiber. To realize a narrow linewidth, the system is cooled down to liquid-helium temperatures. The molecule is optically excited and its fluorescence is collected through the fiber. We have recorded an excitation spectrum, a saturation curve and analyzed the contributions of Raman background fluorescence. This presents to date the crucial limit for the introduced device. The single photon nature is proven by an anti-bunched auto-correlation recording, which also shows coherent Rabi oscillations.
\end{abstract}


\maketitle
\end{CJK*}

\section{Introduction}
Single photon sources are believed to be a key ingredient for quantum communication, quantum information processing, and quantum measurements. The generation of single photons can mainly be achieved by two different approaches: The generation via a parametric process, which allows to generate photons in pairs whereas one is used to herald the presence of its respective partner. Often, these sources are spectrally broad, which renders their interfacing to other quantum system difficult. Furthermore, such sources suffer from multi-photon contributions under high pump powers which ultimately limits the number of available photons, as they need to be pumped weekly enough to decrease these unwanted contributions. An alternative is the generation by single emitters, such as atoms or ions and, in the solid state, quantum dots, defect centers or novel two-dimensional materials. Historically, the first discovered solid-state source of single photons was realized by single molecules, which simultaneously 
 deliver a high flux and a narrow-band emission and hence are still a highly promising system for applications in quantum technology. 

The usability of single photon sources is often enhanced by their single modal generation. For a single photon source to be useful, in most experiments it is necessary to ensure that the photons are generated in a defined spatial mode, ideally matched to the mode of an optical fiber. This allows for convenient fiber coupling to other devices. Also, as in interference experiments, the modal overlap is crucial. A defined mode such as a Gaussian is optimal. Therefore fiber-coupled single photon sources allow for the integration in quantum networks and hence is applicable beyond a restricted lab frame.

In order to facilitate excitation, single solid state emitters have been coupled to optical fibers since their beginning~\cite{orrit_prl_1990}. Still, the ideal device would be fiber coupled in both, excitation and detection. This can be implemented by coupling a single emitter to a fiber end-facet what has been successfully demonstrated for nano diamonds~\cite{schroeder_nl_2011} and quantum dot emitters~\cite{cadeddu_apl_2016}. For coupling efficiently to the end-facet of an optical fiber, as in the field of microscopy, a crucial parameter is the fiber's numerical aperture (NA). For reaching high efficiencies the NA needs to be maximized so that light emitted at large angles is also captured and guided through the utilized fiber.

An alternative way of coupling the emission of solid-state quantum emitters to the guided modes of a fiber is through coupling to the evanescent field of a tapered fiber. This has been demonstrated with a variety of emitters such as quantum dots~\cite{yalla_prl_2012,fujiwara_nl_2011,schell_sr_2015}, nitrogen vacancy centers in diamond~\cite{liebermeister_apl_2014,schroeder_nl_2011,vorobyov_tepjd_2016,fujiwara_ao_2017}, defects in hexagonal boron nitride~\cite{schell_ap_2017}, and also molecules~\cite{skoff_pra_2018}. Nevertheless, coupling to tapered fibers is difficult as the fibers easily get contaminated, break, or lose transmission when being cooled down. All these problems can be avoided by directly coupling to the fiber end-facet. Another alternative would be the integration of single emitters into waveguide structures~\cite{faez_prl_2014,kewes_sr_2016,lombardi_ap_2017,tuerschmann_apa_2019}.

The use of single molecules for single photon generation exhibits a number of advantages, which are crucial for a number of applications in quantum technologies. Here we discuss the properties of single molecules under cryogenic conditions (T$\leq$2~K). On of their advantages is their approximately ten-fold larger T$_1$-time when compared to quantum dots. Also, the amount of spectral diffusion is significantly lower than for other quantum emitters due to their insensitivity against external spin noise, such that Fourier-limited linewidths are commonly observed. Simultaneously, the detected flux of photons reaches up to millions of photons per second while maintaining negligible background contribution.

Here, we show the implementation of a molecule based single photon source implemented by optically coupling a molecule to a high-NA fiber. This configuration is cooled down to liquid-helium temperatures where the molecules' transitions narrow down to their natural lifetime limit. The optical excitation of the molecules is realized by free-space coupling as well as by direct excitation through the fiber. Detection is in all cases accomplished through the fiber which collects the single photons and guides them towards the detection system. 

\begin{figure*}[bth]
  \includegraphics[width=\textwidth]{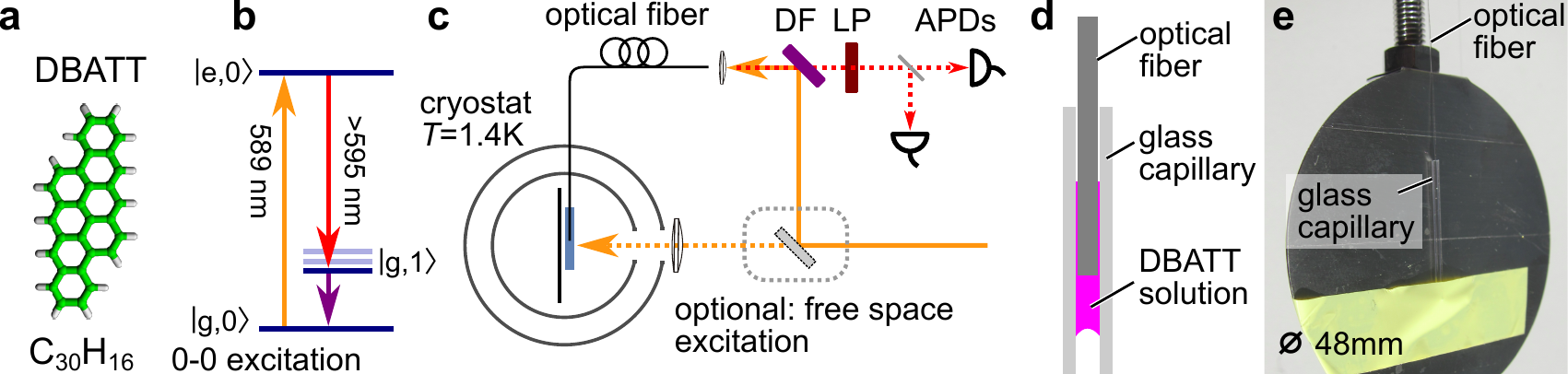}
  \caption{\textbf{Experimental configuration.} a) The 2.3,8.9-dibenzanthanthrene (DBATT) molecule used in this study. b) A simplified level scheme. The resonant zero-phonon-line excitation matches the sodium D$_2$ transition. c) In the experiment, a helium bath cryostat hosts the fiber configuration. The setup allows for the excitation through a cryostat window or through the fiber directly. A Hanbury Brown and Twiss configuration allows for the detection of the photon anti-bunching. DF: dichroic filter; LP: long-pass filter; APDs: avalanche photo diodes. d) For coupling the molecules, the fiber is inserted in a capillary filled with DBATT solution. e) The assembly that is cooled down in the cryostat.}
  \label{fig:fig01}
\end{figure*}

\section{Experiment}
To enhance the collection efficiency of a single molecule in proximity to the fiber end, we have chosen a high-NA optical fiber. Usual single mode glass fibers have a numerical aperture around 0.1--0.13. Here we have used a specialized fiber (UHNA7, Nufern/ Coherent Inc) that is manufactured for a wavelength range of 1500--2000~nm and has a core diameter of 2.4~$\upmu$m while having a very high numerical aperture of 0.41. In the wavelength range used in this experiment, i.e.\ about 600--700~nm, this fiber is not single mode, but the chosen fiber was a compromise between transmission of the fiber and a small core diameter. In another cryogenic setup where shorter fibers can be used, real single mode fibers (such as UHNA440 from Coherent Inc.) can be used, whereas scattering limits the efficiency in long fibers.

The experimental configuration (see Figure~\ref{fig:fig01}) consists of a helium bath cryostat (Janis Inc.), which allows for optical access from the side. The optical fiber is fed into the cryostat through a PTFE-based seal~\cite{abraham_ao_1998}. The overall length of the fiber used is approximately 3~m. 

To excite a single molecule, a single mode dye laser (Coherent 899 dye ring laser, $\Delta\nu\approx$1~MHz) is used. Its linewidth is significantly below the spectral linewidth of the molecules, which enables for resonant excitation and measurement of the molecules' transition linewidths. For the typical fluorescence excitation scheme we use, the utilized filters suppress the laser excitation at the zero-phonon line and allow for the collection of Stokes-shifted photons.

The single molecules are 2.3,8.9-dibenzanthanthrene (DBATT) molecules (CAS: 188-42-1, from W.\ Schmidt, Iggling-Holzhausen, see Fig.~\ref{fig:fig01}a) which are dissolved in a $n$-tetradecane (C$_{14}$H$_{30}$) solution. For realizing the fiber coupled single photon source the solution is soaked into a micro-liter pipette (10~$\upmu$l, Fa.\ Hirschmann). In this way, approx.\ 1-2~$\upmu$l of the solution stays inside the capillary. Subsequently, the stripped and cleaved optical fiber is inserted into the capillary (see also Fig.~\ref{fig:fig01}d). The resulting system is then inserted into the cryostat and cooled down. 

Detection is either realized by two free-space single photon counting modules (SPCM-AQR, Excellitas) in a Hanbury Brown and Twiss configuration or by a compact grating spectrometer (Ocean Optics, QE Pro) for spectrally resolved measurements. The single photon signals are captured by a time-tagger (Swabian Instruments, Time-Tagger 20) which allows for recording of the anti-bunching.

\section{Theory}

\begin{figure}[h!bt]
  \includegraphics[width=\columnwidth]{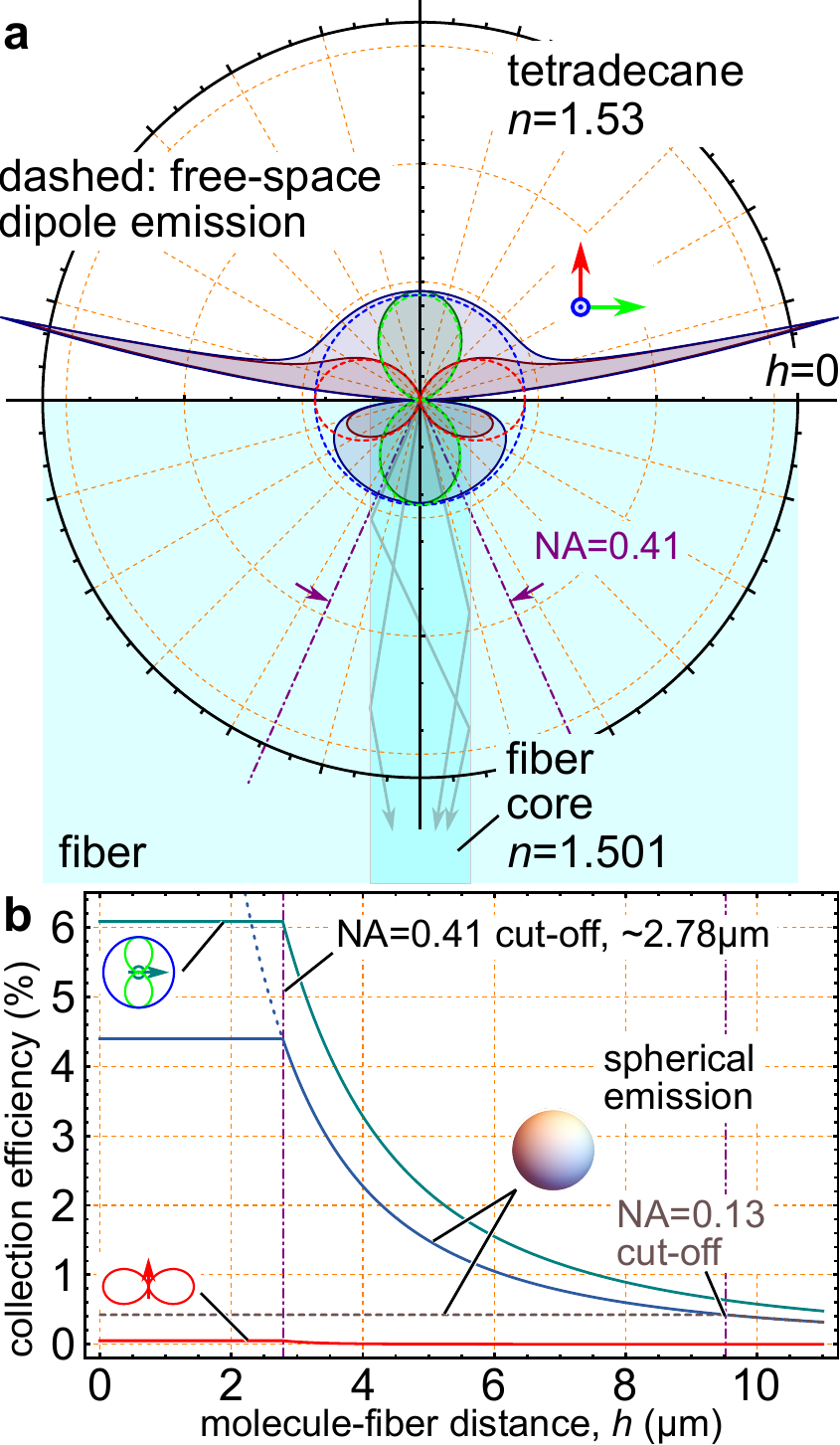}
  \caption{\textbf{Collection efficiency of a molecule into the fiber.} a) Calculated emission pattern of a dipole at a dielectric interface. The refractive indices are given by the molecules' matrix \emph{n}-tetradecane ($n$=1.53) and the fiber core of the utilized glass fiber ($n$=1.501). Subsequently, the emission into the upper hemisphere is favored and for a parallel dipole on the interface around two third of the emission is emitted upwards. b) Full calculation for differently oriented dipoles (red, green). The numerical aperture of the fiber leads to a small distance to the fiber end-facet in which the collection efficiency remains about the same. Blue, gray: calculated collection efficiency assuming spherical emission. The upper dashed curve indicates the pure consideration of the captured solid angle by the fiber core, i.e, setting the NA of the fiber to NA=1. As 50\% of the emission is downwards, this curve reaches 0.5 for a molecule-fiber distance of zero. The lower dashed curve indicates the efficiency for a fiber with NA=0.13.}
  \label{fig:fig02}
\end{figure}

To understand the experimental situation regarding the collection of fluorescent light, other authors use reciprocity theory~\cite{then_pra_2014}. However, here we use another approach and consider an emitting dipole above an optical fiber; the situation is depicted in Fig.~\ref{fig:fig02}a. The numerical aperture of the utilized glass fiber determines how the dipolar emission of the single molecule is captured when the molecule is located directly on top, or in close proximity, to the fiber end-facet. 
Emission into the lower hemisphere on angles that correspond to a NA smaller than the fiber's NA will be captured and guided through the fiber core. Emission outside these angles will not be guided within the fiber core and is lost. If the molecule is further away form the fiber end-facet, geometry determines the amount of light that is captured: Only the solid angle which is formed by the fiber core's geometry, i.e., the effective solid angle at a distance will capture the emitted light. The collection efficiency is therefore reduced with increasing distance between molecule and fiber.

In order to determine the amount of light that meets the above criteria, the effect of the dielectric boundaries in the direct proximity of the dipole has to be taken into account. The light is strongly influenced at such interfaces, as described in the literature~\cite{lukosz_josa_1979}. Here we consider the refractive index of solid \emph{n}-tetradetane, given as $n=$1.53~\cite{morisawa_ao_2017} and the refractive index of the fiber core, which is $n=$1.501. A calculation of the dipolar emission at dielectric interfaces~\cite{lukosz_josa_1979} reveals that 19\% of the emission of an orthogonal dipole and 33\% of the emission of a parallel dipole are emitted into the lower hemisphere, i.e.\ into the direction of the fiber. Fig.~\ref{fig:fig02}a shows the situation for a dipole directly at the interface. The results are shown for an orthogonal dipole (red) and the $r$- and $s$- components of a parallel oriented dipole (blue and green). 

Fig.~\ref{fig:fig02}b shows the collection efficiency of the fiber for a dipole as a function of its distance to the fiber end-facet. Here we use a model, which treats the fiber core as a region with sharp boundaries and considers a single dipole centered on the optical axis. With this simplified model we realize that the numerical aperture of the utilized fiber sets a strict limit to the maximal collection efficiency of the fiber. For a parallel dipole, of which 33\% radiates downwards, this limit is for the given fiber at around 6.1\% of its total radiation. This calculation includes interface effects, the numerical aperture and also the geometry of the problem, as for larger distances not the fibers' NA is the limiting factor, but the solid angle given by the finite size of the fiber core. This limit cuts on at  a distance of 2.78~$\upmu$m and beyond. Below this height, the limitations are dominantly set by the numerical aperture. We like to note, that the redistribution of radiation due to the height of the emitter has essentially no influence within the NA-limited collection cone. This becomes evident in Fig.~\ref{fig:fig02}a, where the lobes captured by the numerical aperture resemble the free-space dipolar pattern for all relevant distances. This is due to the lower index of refraction which is set by the fiber core.

For the situation of an orthogonal oriented dipole, the collection efficiency is drastically reduced by a factor of 117. In the case of \emph{spherical} emission there is no influence by the interface and the refractive indices, such that the collection efficiency is solely set by geometrical parameters. The dashed line which represents the spherical emission in Fig.~\ref{fig:fig02}b would end for a maximal NA of 1.0 at a value of 50\% collection efficiency for a height of zero (dashed line). We like to remark that the geometric cut-off is still found around the distance of 2.78~$\upmu$m from the fiber, which allows to estimate the NA-implied cut-off even with limited mathematical effort. The distance dependence for spherical emission shows a comparable behavior as the full calculation of the orthogonal oriented dipole. This shows that the dominant part of the parallel dipolar emission into the lower hemisphere behaves still proportional to a sphere what implies that the experimental observations of this work are highly likely based on the detection of parallel oriented dipoles, which are located within the first few $\upmu$m above the fiber.

In order to show the influence on the NA of the fiber, the collection efficiency of a standard optical fiber for the visible region of the optical spectrum has also been considered. The commonly used fiber 460-HP (Nufern) has a numerical aperture of 0.13. This leads to the collection efficiency being limited to around 0.5\%, such that experiments with such a fiber will lead to approximately one order of magnitude less light than with the previously discussed fiber. Due to the lower NA, the corresponding cut-off is around 9.5~$\upmu$m, i.e.\ the molecule can be more distant to reach the same effective NA for the collected solid-angle. This is implies that also the detection volume is increased, leading to a higher background in conjunction with a lower collection efficiency -- an unfavorable behavior for our purposes.

We now compare these results to the situation of a usual confocal microscope. In our experimental setup we have numerous examples of very high photon flux and have recorded single molecule emission with more than 1.2~Mio uncorrected detector clicks per second in the past~\cite{siyushev_n_2014,kiefer_apb_2016,rezai_prx_2018}. These results have been acquired by utilizing a solid immersion lens (SIL) with a refractive index of $n=$2.18 of cubic-zirkonia (ZrO$_2$). For a collection lens with an NA of 0.68 (Geltech, C-330) the overall collection efficiency is calculated as 28\%, which is around a factor of 4-5 against the calculated value of fiber collection. Therefore we estimate the count rate to maximally 250~kcps (kilo counts per second), since the rest of the setup remained the same as in previous experiments. With the reductions introduced by spectral filtering (see below), we believe that the results of free-space collection and fiber collection are consistent with each other.

\begin{figure}[thb]
  \includegraphics[width=\columnwidth]{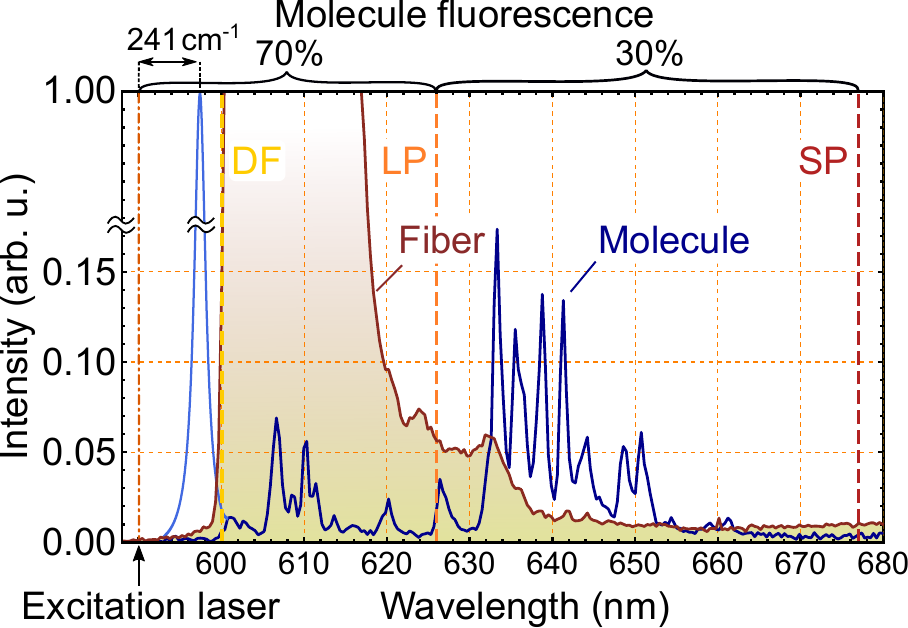}
  \caption{\textbf{Spectral contributions of the experiment.} The utilized optical fiber exhibits a large amount of background contributions until approx.\ 630~nm. Unfortunately, these overlay the majority (about 70\%) of the molecular emission spectrum. In order to remove these unwanted contributions, a long-pass filter around 630~nm has been introduced (LP). Another background contribution of the fiber is above 700~nm, such that a short-pass filter has been introduced (SP). To note: the utilized dichroic filter (DF) we used has a sharp cut-on around 600~nm. Therefore the spectra are only recorded from there on wards.}
  \label{fig:fig03}
\end{figure}

\section{Results}

To understand the emission of the single molecule and also an eventual background contribution of the utilized fibers, the first task was to record a spectrum of the individual components. The spectra of the light emitted by the molecules and of the background light stemming from the fiber are shown in Figure~\ref{fig:fig03}. It can be clearly seen that both contributions have a substantial spectral overlap: Only the higher wavelength side-bands of the molecules are clearly separated from background contributions which stems from the fiber. Hence, only this part can be effectively separated and used in the single photon source.

In order to filter out these unwanted contributions, spectral filtering is employed. A dichroic long-pass filter (cut-on at 600~nm) is used to suppress the excitation laser light. Unfortunately, this filter suppresses the majority of the emission which corresponds to the decay of first electronic excited state to the first excited vibrational state of the electronic ground state. The spectral distance to this band is 241~cm$^{-1}$ from the excitation laser. Still this filtering is not sufficient to filter the remaining background light stemming from the optical fiber. Therefore an additional long-pass with cut-on at 626~nm is used. In order to account for the higher wavelength contributions of the fiber an additional short-pass filter with cut-on at 678~nm is used. The remaining detection window (626~nm--678~nm) contains only about 30\% of the Stokes-shifted photons~\cite{boiron_jcp_1996} emitted by the molecules while the background contributions are suppressed. This reduces the amount of fluorescence photons, but renders the signal-to-background ratio higher, such that single photons can be detected (see below).

The setup can be operated in two configurations: either with free-space excitation of the molecules where the excitation laser enters the cryostat through the windows or by excitation through the fiber. In a first experiment, we employed free-space excitation to reduce the amount of background signal from the fiber. The large focal volume is caused by the $f$=100~mm lens, which focuses down the excitation light to a large spot of around 10-20~$\upmu$m at the fiber tip. The alignment procedure is outlined below. Despite the advantage that the background from the fiber is reduced, the amount of background fluorescence is larger than for the case of excitation through the fiber, due to the large illuminated volume. 

\begin{figure}[t!bh]
  \includegraphics[width=\columnwidth]{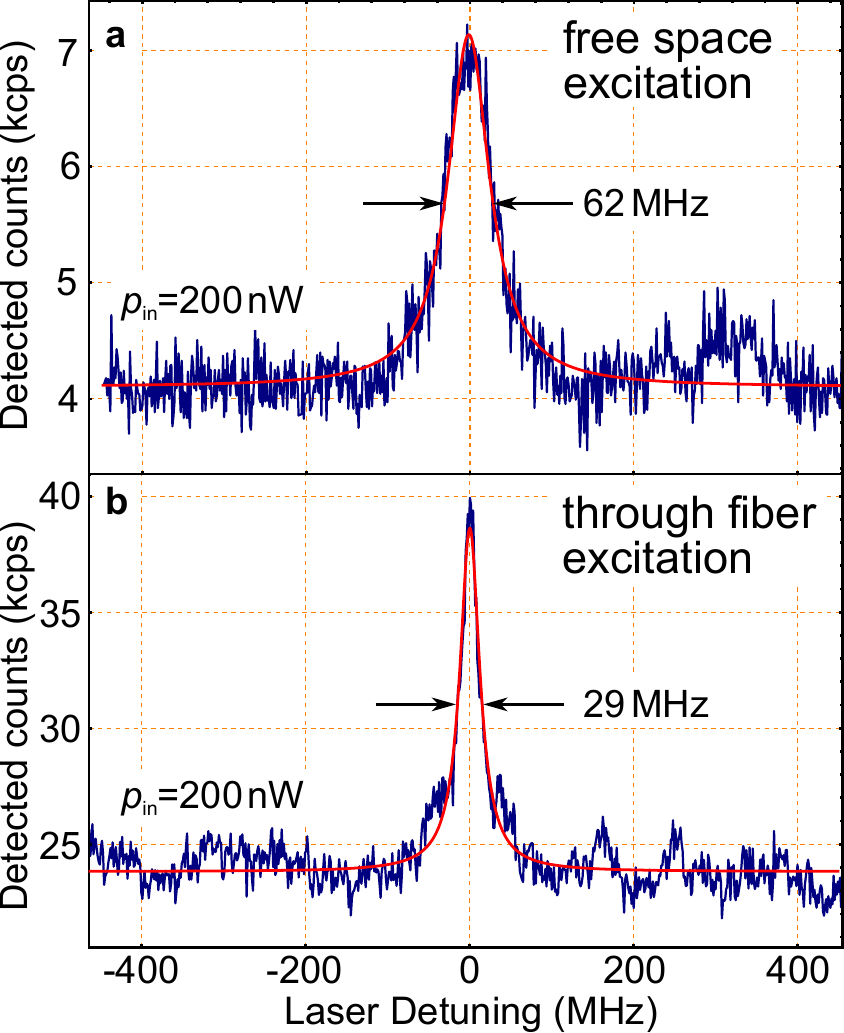}
  \caption{\textbf{Excitation lines of single molecules.} a) Fluorescence excitation spectrum of a single molecule under free-space excitation. The molecule is excited from the side and the fluorescence light is collected through the fiber. b) Line of a single molecule under through-fiber excitation and detection.}
  \label{fig:fig04}
\end{figure}

To enable the free-space excitation at the fiber tip, several mW of excitation light were utilized, and the filters in front of the single photon detector were removed. Simply the scattered light which leaks into the fiber is sufficient to locate the fiber end-facet. Therefore the focus was laterally scanned across the cryogenic configuration by a beam-scanning optics (not shown in Fig.~\ref{fig:fig01}e). When the fiber end was located, single molecule detection was attempted. For this, the filters were re-installed, the excitation laser power was reduced to 200-1000~nW and the laser was frequency scanned through the inhomogeneously broadened band of the molecules around the sodium D-lines. By this excitation scheme we excite the molecules in their zero-phonon line (ZPL) and subsequently detect only Stokes shifted photons. When the excitation laser drives the transition in the zero-phonon line from $\ket{g,0}$ to $\ket{e,0}$ the molecule emits photons into the manifold from $\ket{e,0}$ to $\ket{g,n}$ (see level scheme in Fig.~\ref{fig:fig01}b). Such a ZPL-fluorescence excitation spectrum is shown in Fig.~\ref{fig:fig04}a. With a spectral linewidth of 62~MHz, this particular molecule exceeds the typical linewidth of single DBATT molecules by a factor of 4-5~\cite{kiefer_apb_2016}. Furthermore, the background contribution is large and the signal to background ratio is limited around two, such that a single photon recording would not be applicable. An increase of the laser power does not lead to an enhanced signal-to-background ratio, as commonly observed in single molecule studies.

\begin{figure*}[bth]
  \includegraphics[width=\textwidth]{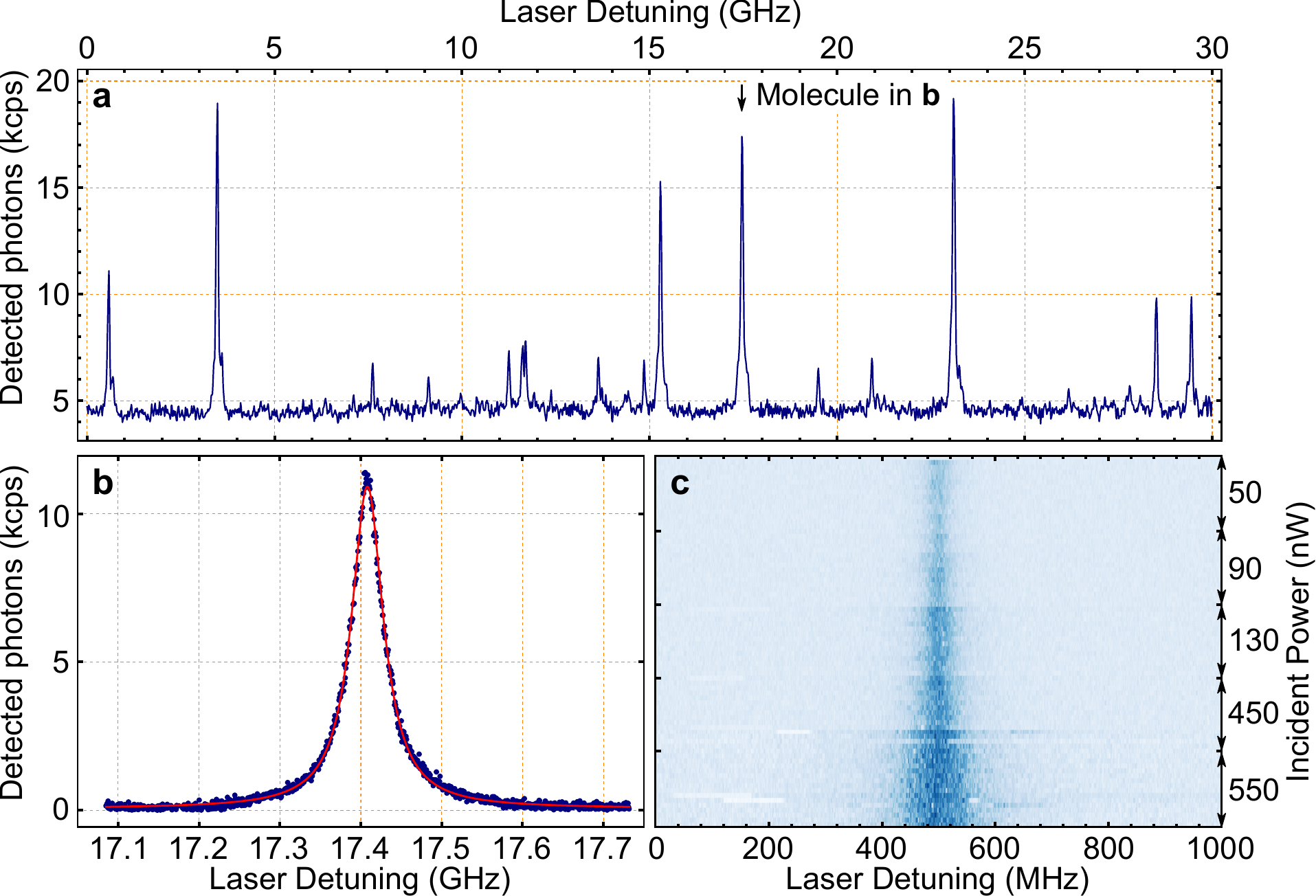}
  \caption{\textbf{Resonant excitation of molecules through the fiber.} a) Fluorescence excitation spectrum of single molecules excited trough the optical fiber with an input power of 50~nW. b) Zoom in into one of the molecules of the spectrum in \textbf{a}. The line can be approximated with a pure Lorentzian. c) A saturation scan where the molecule's line is recorded at different excitation powers.}
  \label{fig:fig05}
\end{figure*}

To further explore the technical possibilities, i.e.,\ to excite a single molecule without the full cryostat setup, we now change to an excitation through the optical fiber directly. Also all alignment steps, except the fiber coupling can be omitted. The excitation of a molecule is shown in Fig.~\ref{fig:fig04}b. While the linewidth is reduced by a factor of two, the signal-to-background ratio is approximately the same. Unlike in the case of free-space excitation, we account the background to a Raman contribution, which appears to be significant. Although the illuminated sample volume is reduced, we do not observe a significantly better signal-to-background ratio.

The above experiments have been conducted on several molecules. Fig.~\ref{fig:fig05} shows an extended resonant excitation measurements of the fiber coupled molecules in the fully fiber coupled configuration. Each peak corresponds to a single molecule that is excited. The emission of the individual molecules resides on a constant level of background photons which stem from the fiber. The heights of the peaks varies, as the coupling efficiency of the molecules is dependent on the exact geometry, i.e., position and orientation relative to the fiber core as outlined in the theory section above.

Each of the detected molecular lines can be further resolved by a finer frequency scan. Fig.~\ref{fig:fig05}b shows a zoom into one of the molecules in the upper panel (labeled). The included fit shows the typical Lorentzian behavior of a life-time limited single emitter. The curve is fitted as:
\begin{equation}
I_{\rm out}(\omega)\propto \frac{1}{(\omega-\omega_0)^2+(\varGamma/2)^2}\, ,
\end{equation}

\noindent
where $\omega_0$ denotes the angular frequency of the transition, $\omega$ the actual angular laser frequency, and $\varGamma$ the linewidth of the emitter.

As the molecules are long-time stable and there is no limitation by photo-bleaching under cryogenic conditions, it is possible to record long measurements at high excitation powers, as is shown in Fig.~\ref{fig:fig05}c, where we present a long-term (10~min) saturation scan. It is evident that the centered line is getting wider and is also increasing in its amplitude when the excitation power is increased, to record a saturation scan, which allows to determine the photo-physical parameters.

\begin{figure}[h!tb]
  \includegraphics[width=\columnwidth]{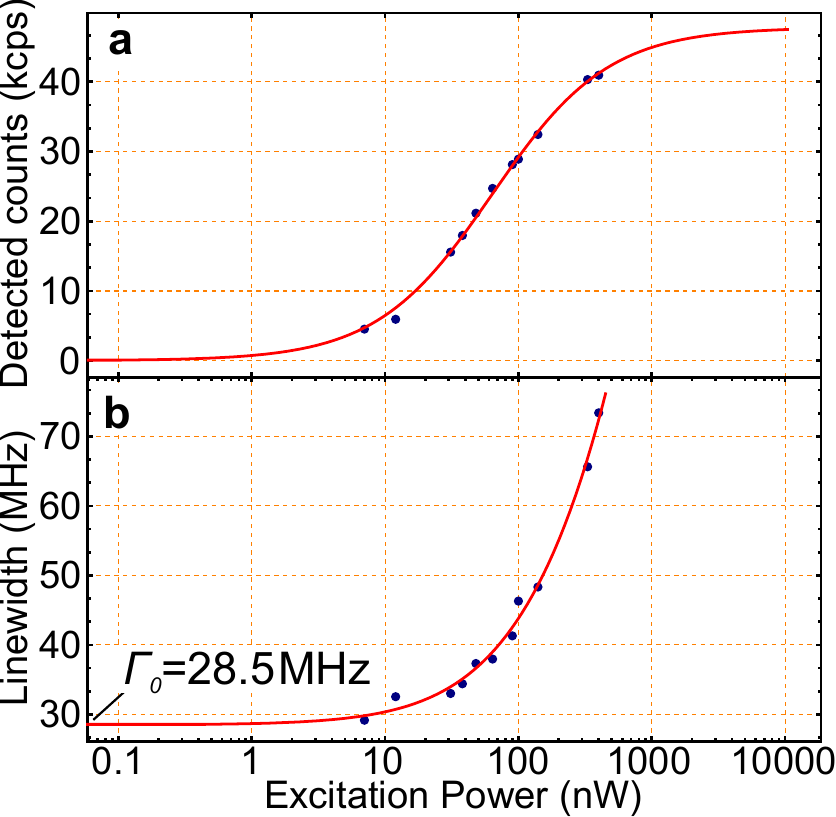}
  \caption{\textbf{Saturation and Linewidth.} a) The saturation curve of the utilized molecule. The behavior corresponds to the usual two-level system broadening behavior. To note: the increasing fiber and impurities background is subtracted in this scan. b) Fitted linewidth of the single molecule under increased excitation intensity. This curve allows for the determination of the natural linewidth, which corresponds to the low-power limit.}
  \label{fig:fig06}
\end{figure}

The saturation scan not only indicates the achievable peak count rates, but gives also an indication how efficient the emitter is excited. The saturation scan is shown in Fig.~\ref{fig:fig06}a, where the emission count rate is shown against the logarithmic incident power into the fiber. It reveals that saturation occurs at 60~nW incident power. This value exceeds the typical ones obtained for the DBATT molecule with a highly focussed excitation laser by a factor of 10-100~\cite{wrigge_np_2008}. The curve is fitted with the equation

\begin{equation}
I_{\rm out}(I_{\rm in})=R_{\infty} \frac{I_{\rm in}}{I_{\rm sat}+I_{\rm in}} \, ,
\end{equation}

\noindent
where $I_{\rm out}$ is the measured number of emitted photons, $I_{\rm in}$ is the excitation intensity, $I_{\rm sat}$ is the saturation intensity and $R_{\infty}$ is the detected count rate at infinite excitation. 

The saturation count rate of around 50~kcps is 10-25 times lower than the achievable count rates in confocal microscopy with a solid immersion lens~\cite{wrigge_np_2008,siyushev_n_2014}. We account this to the reduced coupling efficiency to the fiber, which is outlined in the theory section (factor 4-5). Furthermore it is influenced by a factor which accounts for the reduced spectral contribution of 3-4. To this end the count rates are comparable for the DBATT molecule. With the present background contribution and the still limited NA of the fiber, we believe that the count rates will not be significantly higher until those limitations are addressed.

While the counts of the emitter are saturating, the spectral linewidth in the saturation scan also rises. This is described by the square root proportionality, written as:
\begin{equation}
\varGamma_2 (I_{\rm in})=\varGamma_0 \sqrt{1+I_{\rm in}} \, ,
\end{equation}

\noindent
where the linewidth at the low excitation limit is given as $\varGamma_0$ and the incident intensity as $I_{\rm in}$. For this particular molecule a minimum linewidth of 28.5~MHz is found. This value is larger than the typical values by a factor of approximately two. We account this to eventually opened relaxation channels due to the nearby interface, which may lead to cracks and strain in the formed Shpol'ski\unichar{0301}-matrix. In the course of the experiments no other, narrower, molecules were observed.


\begin{figure}[h!tb]
  \includegraphics[width=\columnwidth]{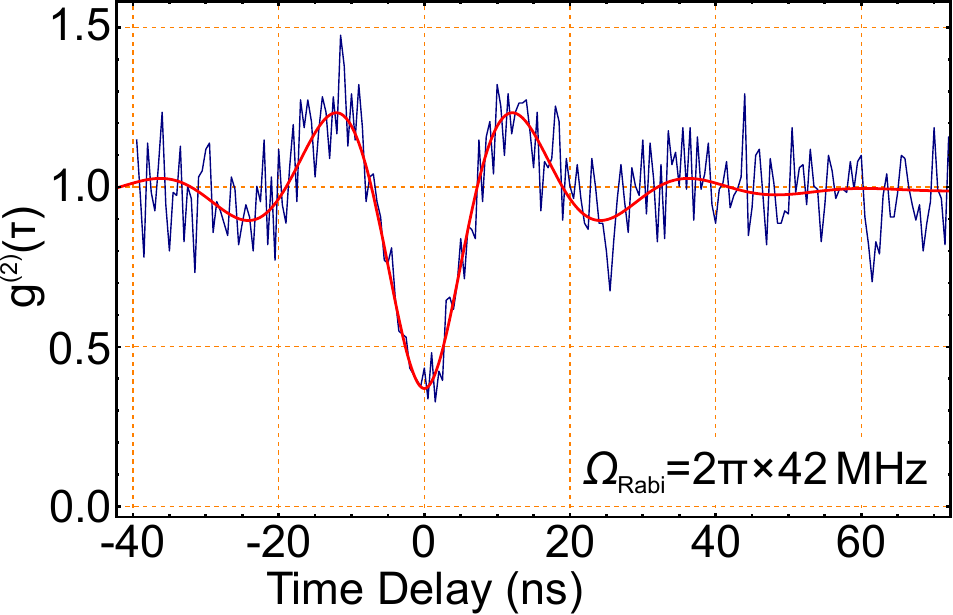}
  \caption{\textbf{Anti-bunching and Rabi oscillations.} The single photon stream is characterized by the photon anti-bunching value for $g^{(2)}(0)$ which is below the value of 0.5 without background correction. Coherent oscillations between the ground and excited state can be seen. The Rabi-frequency is $\varOmega_{\rm Rabi}=2\pi\times 42$~MHz.}
  \label{fig:fig07}
\end{figure}

A crucial value for single photon sources is the measurement of their photon auto-correlation function, $g^{(2)}(\tau)$. A measurement of $g^{(2)}(0)<1$ proves a non-classical behavior of the detected photon stream. Such a behavior can not be obtained for classical light sources. A further reduced value of $g^{(2)}(0)<0.5$ indicates that we are dealing mainly with a single emitter~\cite{rezai_qst_2019}. Fig.~\ref{fig:fig07} shows the photon anti-bunching for the presented fiber-integrated sodium-resonant single photon source. The measurement was performed at an excitation power of 350~nW, and results in an uncorrected $g^{(2)}(0)=0.35\pm0.05$. Together with the measured signal-to-background ratio of 4 in the other measurements, this value could be corrected (see e.g.~\cite{schell_sr_2013}, Eqn.~1) to negligible values of $g^{(2)}(0)$, such that the source can be regarded as a genuine single photon source.

Additionally, the molecules' auto-correlation curve shows coherent oscillations between the ground and the excited state, so-called Rabi oscillations~\cite{wrigge_np_2008,rezai_njp_2019}. To fit the curve in Fig.~\ref{fig:fig07} we use: 
\begin{eqnarray}
g^{(2)}(\tau)=&& \nonumber \\
1-\left(\cos{(\varOmega_{\rm Rabi}|\tau|)}+\frac{\varGamma_1+\varGamma_2}{2 \varOmega_{\rm Rabi}} \cos{(\varOmega_{\rm Rabi}|\tau|)}\right)&& \nonumber \\
\times \, e^{\frac{(\varGamma_1-\varGamma_2)}{2}|\tau|}\, .&& 
\end{eqnarray}

\noindent
We find that $\varGamma_2/\varGamma_1=2$, which indicates a Fourier-limited linewidth of the emitter. As usual, Rabi oscillations further indicate that the molecule is driven above saturation.

\section{Conclusion \& Outlook}

In conclusion, we have presented a fiber-integrated single photon source resonant to the sodium D$_2$ transition based on single DBATT molecules. The approach used to build the source is very general and can be applied to different molecules or even completely different emitters. Therefore a change of the wavelength is straight forward. For example, it is possible to use the molecule dibenzoterrylene (DBT) in order to do reach to the atomic rubidium or potassium transitions~\cite{siyushev_n_2014}. Considering that Raman scattering scales with the fourth power, this would reduce the Raman scattering in the utilized optical fiber by a factor of approximately three. In the current implementation, the collection efficiency of the source is limited due to spectral filtering and the presence of the high refractive index $n$-tetradecane in which the molecules are hosted. Therefore, to further increase the efficiency, a shorter fiber or one with less background can be employed, which relaxes the need for spectral filtering of the Raman contribution. Alternatively, a resonant excitation scheme can be implemented, such that the molecules are excited on their zero phonon line, and the detection is realized in another polarization mode on the same wavelength. Moreover it would be possible to remove the high refractive index $n$-tetradecane solution, e.g., by placing a thin, sublimated crystal hosting the molecules on the fiber end-facet~\cite{pfab_cpl_2004,polisseni_oe_2016}. Then the coupling efficiency of the single molecule into the fiber would be increased.

The reported source can be an important resource for quantum technology and reduces the experimental complexity of cryogenic single molecule studies. We expect that a fully fiber integrated source like the one presented here can be immediately used in applications such as quantum sensing and quantum key distribution.

\section*{Acknowledgments}
We acknowledge funding from the Deutsche Forschungsgemeinschaft in the project GE2737/5-1, the Max Planck Society, and the COST Action MP1403 ``Nanoscale Quantum Optics'' funded by COST (European Cooperation in Science and Technology). We also acknowledge discussions with Dr.\ Bert Hecht, W\"urzburg.
\end{document}